\documentclass[12pt,twoside]{article}   %
\usepackage[super]{cite}
\usepackage{amsmath,amssymb,amsfonts}
\usepackage{booktabs}
\usepackage{algorithmic}
\usepackage{graphicx}
\usepackage{textcomp}
\usepackage{tikz}
\usepackage{rotating}
\usetikzlibrary{positioning}
\usepackage{multirow} %
\usepackage{makecell}
\usepackage{soul}
\usepackage{subcaption} %

\usepackage{fancyhdr}		%

\usepackage[section]{placeins}   %

\usepackage{graphicx}

\makeatletter \renewcommand\@biblabel[1]{$^{#1}$} \makeatother
\newif\ifblind
\blindfalse   %

\ifblind
\else
\fi

\usepackage{hyperref}
\hypersetup{ colorlinks,
    citecolor=blue,
    filecolor=blue,
    linkcolor=blue,
    urlcolor=blue
}

\definecolor{gray}{rgb}{0.6,0.6,0.6}
\definecolor{red}{rgb}{0.85,0,0}
\definecolor{green}{rgb}{0,0.85,0}
\definecolor{blue}{rgb}{0,0,0.85}
\definecolor{beige}{rgb}{0.92,0.87,0.78}
\usepackage[all]{hypcap}    %

\begin{document}

\begin{center}
\sf
{\Large \bfseries Comparison of Deep Learning and Particle Smoother EM Methods for Estimation of Rb-82 Myocardial Perfusion PET Kinetic Parameters} \\  
\vspace*{10mm}

\ifblind
[Author information will be provided after peer review]
\else
Myungheon Chin$^{1,2,*}$, Sarah J. Zou$^{1,2,*}$, Garry Chinn$^2$, and Craig S. Levin$^{1\text{--}3}$ \\
$^1$ Department of Electrical Engineering, Stanford University\\
$^2$ Department of Radiology, Stanford University\\
$^3$ Departments of Bioengineering and Physics and Molecular Imaging Program at Stanford (MIPS), Stanford University \\
$^*$ These authors contributed equally to this work.
\fi

\vspace{5mm}
Version typeset \today
\end{center}

\pagenumbering{roman}
\setcounter{page}{1}
\pagestyle{plain}

\ifblind

\else
Author to whom correspondence should be addressed. email: cslevin@stanford.edu \\
\fi

\begin{abstract}
\noindent {\bf Background:} Positron emission tomography (PET) enables quantification of dynamic physiological processes through time-resolved imaging. In $^{82}$Rb myocardial perfusion PET, kinetic compartment modeling is used to estimate physiological parameters and derive myocardial blood flow. However, conventional nonlinear least squares (NLLS) estimation is sensitive to model misspecification when not all parameters can be reliably estimated and must instead be fixed or initialized using population averages, which can degrade accuracy.\\
{\bf Purpose:} This work develops and evaluates two alternative kinetic analysis approaches for $^{82}$Rb PET: a particle smoother-based Expectation--Maximization method (PSEM) and a convolutional neural network (CNN).\\
{\bf Methods:} Both methods were evaluated using simulated $^{82}$Rb dynamic myocardial perfusion studies and compared against NLLS and a Kalman smoother-based Expectation--Maximization (KEM) algorithm across multiple frame durations and noise levels.\\
{\bf Results:} Across 2--10~s frames, the CNN achieved the lowest relative errors for all parameters ($F$: 8.78--4.98\%, $k_3$: 26.05--25.50\%, $k_4$: 34.34--22.76\%), significantly outperforming NLLS, KEM, and PSEM (Holm-adjusted $p < 10^{-15}$ at 1.0$\times$ noise, 2~s frames), although performance degraded under out-of-distribution input-function conditions.\\
{\bf Conclusions:} Overall, the CNN provided the most accurate and robust in-distribution kinetic parameter estimates across frame durations. In contrast, PSEM exhibited parameter-dependent behavior, improving $k_3$ estimation while underperforming for $F$, suggesting that further methodological refinement is needed.
\end{abstract}

\newpage     %

\tableofcontents

\newpage

\setlength{\baselineskip}{0.7cm}      %

\pagenumbering{arabic}
\setcounter{page}{1}
\pagestyle{fancy}
\section{Introduction}
\label{sec:introduction}
Positron emission tomography (PET) is a nuclear medicine imaging technique that measures the three-dimensional distribution of radiolabeled molecules, known as tracers. Dynamic PET uses time-resolved imaging data to produce time activity curves (TAC) for image voxels or specified regions of interest (ROIs) to estimate kinetic parameters such as cardiac flow and receptor binding rate constants. \cite{bailey_springer_2005}. This paper examines the kinetic analysis of dynamic myocardial perfusion PET studies conducted with the tracer $^{82}$Rb.

Cardiovascular disease remains one of the leading causes of death worldwide, making accurate risk assessment and early identification of vulnerable patients critically important \cite{naghavi_cir_2003}. Dynamic positron emission tomography (PET) enables noninvasive imaging of the temporal dynamics of molecular processes \textit{in vivo} \cite{bailey_springer_2005} and provides functional information for the assessment of coronary artery disease (CAD) through quantification of regional myocardial blood flow (MBF) \cite{huang_ajphcp_1989, schwaiger_ajc_1991, herrero_cr_1992, lortie_ejnmmi_2007, yoshinaga_jc_2010, sohn_jti_2022}. Several radiotracers are commonly used for PET-based MBF quantification, including $^{15}$O-water, $^{13}$N-ammonia, $^{82}$Rb, $^{11}$C-acetate, and $^{18}$F-flurpiridaz \cite{schwaiger_ajc_1991, sohn_jti_2022}. Among these, $^{82}$Rb is widely adopted in clinical practice because it is produced using an $^{82}$Sr\:/\:$^{82}$Rb generator and has a short half-life of 76.4~s, enabling multiple perfusion measurements within a single imaging session \cite{yoshinaga_jc_2010, sohn_jti_2022}. However, the short half-life necessitates rapid acquisition and leads to reduced signal-to-noise ratio. $^{82}$Rb exhibits a relatively low and flow-dependent myocardial extraction fraction, resulting in a nonlinear relationship between tracer uptake and true MBF, particularly at higher flow rates and  short acquisition frames further limit sensitivity to slower kinetic components \cite{chatal_story_2015} . By comparison, $^{13}$N-ammonia offers higher and more stable extraction and a shorter positron range, yielding improved signal-to-noise ratio and more linear flow sensitivity, albeit requiring cyclotron production, whereas $^{15}$O-water is an ideal freely diffusible tracer but presents logistical and modeling challenges due to its extremely short half-life \cite{murthy_clinical_2018}. 

Quantification of MBF via dynamic PET imaging requires appropriate kinetic modeling \cite{gunn_jcbfm_2001, morris_et_2004, watabe_anm_2006}. The radioactive isotope $^{82}$Rb is typically modeled using a two-tissue compartmental model \cite{huang_ajphcp_1989, herrero_cr_1992, yoshinaga_jc_2010}. Estimating kinetic parameters in multi-tissue compartment scenarios often involves optimization challenges \cite{gunn_jcbfm_2001, morris_et_2004, bailey_springer_2005, watabe_anm_2006}, as these objective functions are not well conditioned and can have multiple local minima, complicating parameter estimation. Standard approaches like nonlinear least squares (NLLS) and graphical methods, such as the Patlak or Logan plots \cite{patlak_jcbfm_1983, logan_jcbfm_1990, watabe_anm_2006} differ primarily in their underlying assumptions and intended use cases. NLLS methods, though computationally demanding, offer estimation of micro-parameters at the cost of increased sensitivity to model assumptions, particularly in dynamic, low-count settings.  Graphical approaches provide robust and precise estimates when their assumptions (e.g., irreversible trapping or equilibrium conditions) are satisfied and provide macro-parameter estimation. Graphical approaches are widely used for whole-body/total-body kinetic modeling \cite{karakatsanis_dynamic_2013, wu_whole-body_2022}.  

In our study, we investigated two novel methods for estimating PET kinetic parameters by simulating a two-tissue compartmental model of $^{82}$Rb and assessing their performance against the trust region-based NLLS method and a Kalman-smoother EM (KEM) benchmark. The KEM, similar to what Gibson and Ninnes described \cite{gibson_automatica_2005}, leverages the expectation-maximization (EM) algorithm, which iteratively alternates between estimating hidden states using the Kalman smoother \cite{kalman_jbe_1960, rauch_aiaa_1965, roweis_nc_1999} (E-step) and maximizing the Q function (M-step). The EM algorithm decomposes the multi-dimensional parameter solution space into a more manageable lower-dimensional optimization problem. The first novel method, the particle-smoother EM (PSEM), extends the EM algorithm by replacing the Kalman smoother with a particle smoother \cite{gordon_ieepf_1993, schon_automatica_2011}. This provides greater flexibility as Kalman smoother requires unimodal Gaussian posterior distributions and the particle smoother allows for multimodal parameter distributions and nonlinear constraints. We hypothesize these properties could be advantageous under low-count, early-frame conditions or for more complex kinetic models.

The second novel approach involves a convolutional neural network (CNN), a deep learning strategy that exploits spatial and temporal coherencies in datasets \cite{lecun_nc_1989, alzubaidi_jbd_2021}. CNNs have shown strong performance in time-series analysis \cite{gamboa_arxiv_2017} and PET applications \cite{wang_mp_2021, hu_nssmic_2019}. To our knowledge, this study represents the first application of a CNN to dynamic $^{82}$Rb PET. We enhanced a traditional CNN model by incorporating a Time2Vec layer \cite{kazemi_arxiv_2019} to better capture temporal information. The CNN approach makes no assumptions about the underlying compartment model, allowing further flexibility in finding a mapping between PET activity and kinetic parameters. These methods were compared to assess their accuracy and efficiency in estimating PET kinetic parameters.

\section{Methods} \label{sec:methods}
\subsection{2TC Model for $^{82}$Rb PET}
\begin{figure}[h]
\centering
\begin{tikzpicture}[node distance=3cm, auto]
\node [draw, rectangle, minimum width=1.5cm, minimum height=1cm] (A) {$C_a(t)$};
\node [draw, rectangle, minimum width=1.5cm, minimum height=1cm, right = 1cm of A, align=center] (B) {$q_1(t)$\\Fast\\Exchangeable};
\node [draw, rectangle, minimum width=1.5cm, minimum height=1cm, right = 1cm of B, align=center] (C) {$q_2(t)$\\Slow\\Exchangeable};
\draw [->, shorten >=2pt, shorten <=2pt] (A) to [out=30,in=150] node[midway, above] {$F$} (B);
\draw [->, shorten >=2pt, shorten <=2pt] (B) to [out=30,in=150] node[midway, above] {$k_{3}$} (C);
\draw [->, shorten >=2pt, shorten <=2pt] (C) to [out=210,in=330] node[midway, below] {$k_{4}$} (B);
\draw [->, shorten >=2pt, shorten <=2pt] (B) to [out=210,in=330] node[midway, below] {$F/V1$} (A);
\end{tikzpicture}
\caption{2TC Model for $^{82}$Rb: The inflow and outflow rates between the plasma and the fast exchangeable compartment are assumed to be equal ($F$), but the flow rate from the first state to the plasma is divided by $V1$ since $q_1(t)$ is not a concentration but the total amount of the activity.}
\label{fig:rb82_model}
\end{figure}
Regional myocardial perfusion PET with \(^{82}\)Rb is well modeled by a two-compartment model \cite{mullani_myocardial_1983, huang_ajphcp_1989, herrero_cr_1992}. In Fig.~\ref{fig:rb82_model}, \(F\) is flow from plasma $C_a(t)$ to the first compartment $q_1$, \(V1\) is its volume (with \(V1=v\) times tissue volume), and $k_3$/$k_4$ are exchange rates with the second compartment $q_2$. The model is:
\begin{subequations}\label{eq:2TC_Rb}
\begin{equation}
    \frac{\mathrm{d} q_1(t)}{\mathrm{d}t} =F C_a(t) - (\frac{F}{V1} + k_3) q_1(t) + k_4 q_2(t) 
\end{equation}
\begin{equation}
    \frac{\mathrm{d} q_2(t)}{\mathrm{d}t} = k_3 q_1(t) - k_4 q_2(t) 
\end{equation}
\end{subequations}

where $C_a(t)$ is the plasma input and $q_1(t), q_2(t)$ are compartment activities. In state-space form,
\begin{subequations}\label{eq:gen_state}
\begin{equation}
    x(t) = \
    \begin{bmatrix}
    q_1(t) \\
    q_2(t)
    \end{bmatrix}, 
    u(t) = \
    \begin{bmatrix}
    C_a(t) 
    \end{bmatrix} 
\end{equation}
\begin{equation}
\label{eq:AB}
     A = \begin{bmatrix}
     - (F/V1 + k_3) & k_4 \\
    k_3 & - k_4
    \end{bmatrix}, 
    B =    
    \begin{bmatrix}
     F\\ 
     0
    \end{bmatrix}
\end{equation}
\begin{equation} \label{eq:x}
    \dot{x} = 
    A
    x + 
    B
    u
\end{equation}
\begin{equation} \label{eq:y}
    y = Cx + Du
\end{equation}
\end{subequations}
The output $y$ is the decay-corrected ROI activity. For $^{82}$Rb,

\begin{equation} \label{eq:CD_rb82}
    C = 
    \begin{bmatrix}
        1 - F_{P} & 1 - F_{P}
    \end{bmatrix}, 
    D = 
    \begin{bmatrix}
        F_{P}
    \end{bmatrix}
\end{equation}
where $F_{P}$ represents the plasma fraction of tracer concentration in ROI.

\begin{equation} \label{eq:state_sol}
    x(t) = e^{A (t - t_0)}x(t_0) + \int_{t_0}^t  e^{A (t - \tau)} Bu(\tau) d\tau
\end{equation}

We discretize at sampling interval \(T\) using matrices \(G\) and \(H\):

\begin{equation} \label{eq:forward}
   x((n + 1)T) = G(T)x(nT) + H(T)u(nT)
\end{equation}
\begin{equation} \label{eq:measurement}
   y(nT) = Cx(nT) + Du(nT)
\end{equation}

Assuming zero-order hold, the discrete-time state update is
\begin{equation}
x((n + 1)T) = e^{AT}x(nT) + \int_{nT}^{(n+1)T} e^{A[(n+1)T-\tau]} B \, u(nT) \, d\tau .
\end{equation}

Comparing with \eqref{eq:forward} gives
\begin{subequations} \label{eq:GH}
\begin{equation}
G(T) = e^{AT},
\end{equation}
\begin{equation}
H(T) = \int_{0}^{T} e^{A\lambda} B \, d\lambda .
\end{equation}    
\end{subequations}

Here, $\lambda = (n + 1)T - \tau$. The discrete state-space model is a hidden Markov model; the hidden states $x$ serve as EM latent variables.

PET measurements are frame-averaged over $[T_a, T_b]$ with decay constant $\kappa$:

\begin{subequations} \label{eq:measurements}
\begin{equation} \label{eq:measurement_PET}
    y_{PET} = \int_{T_a}^{T_b}  e^{- \kappa t} y(t) \, dt
\end{equation}
\begin{equation} \label{eq:measurement_input}
    u_{PET} = \int_{T_a}^{T_b}  e^{- \kappa t} u(t) \, dt
\end{equation}
\end{subequations}

\subsection{Data Generation}
Kinetic parameters were sampled uniformly within ranges in Table~\ref{tab:sim_params} \cite{herrero_cr_1992}. Total tissue volume was 10~mL, and the initial state was zero. The default input function followed Herrero et al.~\cite{herrero_cr_1992}, \( C_a(t) = \frac{a t^4}{t^5 + b} \), with $a$ and $b$ sampled uniformly within 90--110\% of the reported values (39{,}218 and 1{,}428). The input function drove \eqref{eq:2TC_Rb} to generate state trajectories. PET measurement noise was added using a VMR model based on frame duration (Appendix~\ref{sec:noise_estimation}).

\begin{table}[h]
\caption{The range of each kinetic parameter used in the simulations.}
    \centering
    \begin{tabular}{|c|c|c|c|}
        \hline
        \textbf{Parameter} & \textbf{Units} &\textbf{Lower Bound} & \textbf{Upper Bound} \\
        \hline
        $F$ & $\text{mL}\ \text{s}^{-1}$& $0.00167$ & $0.0667$ \\
        $k_3$ & $\text{ s}^{-1}$ & $0.00167$ & $0.0667$ \\
        $k_4$ & $\text{ s}^{-1}$ & $0.000167$ & $0.01667$ \\
        $v$ & {unitless} & 0.1 & 0.9 \\
        $F_{p}$ & {unitless} & 0.1 & 0.9 \\
        \hline
    \end{tabular}
    
    \label{tab:sim_params}
\end{table}

After simulating $y_{\text{PET}}[n]$ with 2, 5, and 10~s frames, we applied decay correction and linear interpolation to obtain $\hat{y}$ on a 0.5~s grid using \texttt{scipy.interpolate.interp1d} (\texttt{bounds\_error=False}, \texttt{fill\_value="extrapolate"}). The interpolated curves were cropped to 1{,}024 samples (512~s), and $\hat{u}$ was generated similarly. The same pipeline was applied to all methods. We generated 80{,}000 training/validation datasets, 200 test datasets, and 400 out-of-distribution (OOD) datasets (200 step-function, 200 gamma-variate with multi-exponential washout; Sec.~\ref{sec:ood_func}); sets are disjoint.

\subsubsection{Out of Distribution Generation} \label{sec:ood_func}
We used two OOD arterial input functions (AIFs):

\begin{enumerate}
    \item \textbf{Step--function AIF}
    \[
        C_{\mathrm{step}}(t) = 
        \begin{cases}
            H, & 0 \le t \le T_{\mathrm{dur}}, \\[4pt]
            0, & t > T_{\mathrm{dur}},
        \end{cases}
    \]
    where $H$ is the input amplitude and $T_{\mathrm{dur}}$ is the pulse duration.

    \item \textbf{Gamma--variate bolus with multi--exponential washout (OOD-AIF)}
    \[
        C_{\mathrm{gamma}}(t)
        = A_{\mathrm{b}} \, 
          t^{\alpha} e^{-t/\beta}
        \;+\;
        A_{1} e^{-\lambda_{1} t}
        \;+\;
        A_{2} e^{-\lambda_{2} t}
        \;+\;
        A_{3} e^{-\lambda_{3} t},
    \]
    where the first term models the bolus peak (shape $(\alpha,\beta)$, amplitude $A_{\mathrm{b}}$) and the exponential terms model washout with rates $\lambda_{1},\lambda_{2},\lambda_{3}$.
\end{enumerate}

\subsection{Parameter estimation algorithms}
\subsubsection{Nonlinear least squares (NLLS)}

We used a trust-region nonlinear least squares (NLLS) baseline to estimate the kinetic parameters. Given parameters $\theta$ and residuals $f_i(\theta)$, the objective function is

\begin{equation}
\label{eq:nlls}
\Phi(\theta) = \frac{1}{2} \lVert f(\theta) \rVert^2
= \frac{1}{2} \sum_{i=1}^{n} f_i(\theta)^2.
\end{equation}

The optimization was performed using the trust-region solver implemented in SciPy \cite{scipy}, without additional regularization. In the main experiments, a single initialization was used. To assess sensitivity to initialization, a retrospective analysis with multiple starting points is provided in Appendix~\ref{sec:nlls_init}.

A maximum of 10 iterations was allowed (Table~\ref{tab:estimation_settings}). The solver could terminate earlier if standard convergence criteria were satisfied, including small relative change in the objective function (\texttt{ftol}), small parameter update norm (\texttt{xtol}), or small gradient norm (\texttt{gtol}), using default SciPy tolerances when not explicitly specified. This configuration was applied consistently across all NLLS experiments.

\subsubsection{EM algorithm} \label{sec:k-em}

The EM algorithm performs iterative optimization by alternating between an Expectation (E) step and a Maximization (M) step. In the E-step, the log-likelihood is lower-bounded using Jensen’s inequality by evaluating \( \mathcal{F} \) in \eqref{eq:jensen} and defining an auxiliary distribution \( Q \). Specifically, the E-step updates $Q_{k+1} = p_{\theta_k}(x_{1:N}\mid y_{1:N})$, after which the M-step maximizes \( \mathcal{F} \) with respect to \( \theta \) \cite{roweis_nc_1999}.

With latent states \(x_{1:N}\) and measurements \(y_{1:N}\), the log-likelihood is

\begin{equation}
    \mathcal{L}(\theta) = \log p_{\theta}(y_{1:N})
    = \log\int p_{\theta}(x_{1:N}, y_{1:N})\,dx_{1:N}.
\end{equation}

Introducing an auxiliary distribution \(Q(x_{1:N})\) and applying Jensen’s inequality yields the lower bound

\begin{subequations}
\label{eq:jensen}
\begin{equation}
\log \int p_{\theta}(x_{1:N}, y_{1:N})\,dx_{1:N}
= \log\int Q(x_{1:N})\frac{p_{\theta}(x_{1:N}, y_{1:N})}{Q(x_{1:N})}\,dx_{1:N}
\end{equation}
\begin{equation}
\geq \int Q(x_{1:N})\log\frac{p_{\theta}(x_{1:N}, y_{1:N})}{Q(x_{1:N})}\,dx_{1:N}
\end{equation}
\begin{equation} \label{eq:QV}
= \int Q(x_{1:N}) \log p_{\theta}(x_{1:N}, y_{1:N})\,dx_{1:N}
- \int Q(x_{1:N}) \log Q(x_{1:N})\,dx_{1:N}
\end{equation}
\begin{equation}
= \mathcal{F}(Q,\theta).
\end{equation}
\end{subequations}

\textbf{EM Algorithm}
\begin{subequations}
\begin{equation}
    \text{E-step}: \quad Q_{k+1} \leftarrow p_{\theta_{k}}(x_{1:N}| y_{1:N})
\end{equation}
\begin{equation} \label{eq:M}
    \text{M-step}: \quad 
    \theta_{k+1} \leftarrow 
    \operatorname*{argmax}_{\theta}
    \int p_{\theta_{k}}(x_{1:N}|y_{1:N})
    \log p_{\theta}(x_{1:N}, y_{1:N})\,dx_{1:N}.
\end{equation}
\end{subequations}

The M-step objective can be decomposed as \cite{schon_automatica_2011}:

\begin{equation}
\int p_{\theta_{k}}(x_{1:N} | y_{1:N}) 
\log p_{\theta}(x_{1:N}, y_{1:N}) \,dx_{1:N}
= I_1 + I_2 + I_3,
\end{equation}

where

\begin{subequations} \label{eq:Is}
\begin{equation}
I_1 = \int \log p_{\theta}(x_1)\, p_{\theta_k}(x_1 | y_{1:N}) \,dx_1,
\end{equation}
\begin{equation}
I_2 = \sum_{t=1}^{N-1} 
\iint \log p_{\theta}(x_{t+1}|x_t)\,
p_{\theta_k}(x_t, x_{t+1} | y_{1:N})\, dx_t dx_{t+1},
\end{equation}
\begin{equation}
I_3 = \sum_{t=1}^{N} 
\int \log p_{\theta}(y_{1:t}|x_t)\,
p_{\theta_k}(x_t | y_{1:N})\, dx_t.
\end{equation}
\end{subequations}

Assuming a linear dynamical system with Gaussian noise, the E-step uses a Kalman smoother \cite{shumway_springer_2000}; implementation details for KEM are in \cite{roweis_nc_1999, gibson_automatica_2005}.

\subsubsection{PSEM}
We briefly describe the particle filter and particle smoother here and show that the three terms in \eqref{eq:Is} can be calculated with the particle smoother-based approach. A particle filter is also known as a sequential Monte Carlo method and can be used to find the solution of non-convex problems. It uses multiple ``particles'' to sample the solution space at each iteration~\cite{schon_automatica_2011, elfring_sensors_2021}.

The particle filter approximates the posterior with weighted samples:

\begin{equation} \label{eq:sum_of_weights}
p(x_{1:k} | y_{1:k}) \approx \sum_{i=1}^{N_s} w_{k}^i \delta(x_{1:k} - x_{1:k}^{i})
\end{equation}

Here \(x_{1:k}^i\) is the $i$th particle trajectory, \(w_k^i\) its weight, and $N_s$ the number of particles. Weights are

\begin{equation} \label{eq:weight}
w_k^i \propto \frac{p(x_{1:k}^i | y_{1:k})}{q(x_{1:k}^i | y_{1:k})}
\end{equation}

where $q$ is the importance density \cite{elfring_sensors_2021}; the recursive update is \cite{arulampalam_tsp_2002, elfring_sensors_2021}:

\begin{equation} \label{eq:weight_update}
w_k^i \propto w_{k-1}^i \frac{p(y_k | x_k^i)p(x_k^i | x_{k-1}^i)}{q(x_k^i | x_{k-1}^i, y_k)}
\end{equation}

Thus the posterior can be approximated as \cite{arulampalam_tsp_2002, elfring_sensors_2021}:

\begin{equation} \label{eq:pf_posterior}
p(x_k | y_{1:k}) \approx \sum_{i=1}^{N_s} w_k^i \delta(x_k - x_k^i)
\end{equation}

We use the prior as the importance density \cite{gustafsson_aesm_2010}:

\begin{equation}
q(x_k | x_{k-1}^{i}, y_k) = p(x_k | x_{k-1}^{i})
\end{equation}

This reduces the weight update equation \eqref{eq:weight_update} to the following expression:

\begin{equation} \label{eq:weight_update_with_prior}
w_k^i \propto w_{k-1}^i p(y_k | x_k^i)
\end{equation}

The particle filter algorithm is:

\textbf{Particle Filter Algorithm}

For each $k$th iteration with input $\{x_{k-1}^i, w_{k-1}^i\}_{i=1}^{N_s}, y_k$, perform the following steps to calculate $\{x_k^i, w_k^i\}_{i=1}^{N_s}$. 
\begin{enumerate}
    \item For $i = 1$ to $N_s$,
    \begin{enumerate}
        \item Forward particle by $x_k^i \sim q(x_k^i | x_{k-1}^i, y_k)$
        \item Update weights $w_k^i$ with \eqref{eq:weight_update_with_prior}
    \end{enumerate}
    \item Perform resampling if necessary.$^{*}$
\end{enumerate}
($^{*}$: Resampling, which is an essential element of the particle filter, is the most common and effective way of preventing particle degeneration \cite{elfring_sensors_2021, schon_automatica_2011, elfring_sensors_2021}.)

The particle smoother estimates the smoothed density \(p_{\theta}(x_t | y_{1:N})\) rather than the filtered density \(p_{\theta}(x_t | y_{1:t})\). The following relationship holds \cite{schon_automatica_2011}:

\begin{subequations} \label{eq:ps}
    \begin{equation}
    p_{\theta}(x_t | y_{1:N}) \approx \sum_{i=1}^{N_s} w_{t|N}^{i} \delta(x_t - x_{t}^{i}),
    \end{equation}
    \begin{equation} \label{eq:smoothed_weight}
    w_{t|N}^{i} = \frac{w_{t}^{i}}{\sum_{k=1}^{N_s} w_{t+1|N}^{k} \frac{p_{\theta}(x_{t+1}^{k}|x_{t}^{i})}{v_t}},
    \end{equation}
    \begin{equation}
    v_t^{k} \triangleq \sum_{i=1}^{N_s} w_{t}^{i} p_{\theta}(x_{t+1}^{k}|x_{t}^{i}).
    \end{equation}
\end{subequations}

Using \eqref{eq:smoothed_weight}, the particle smoother is executed as follows \cite{schon_automatica_2011}:

\textbf{Particle Smoother Algorithm}

\begin{enumerate}
    \item Compute particles \(\{x_t^i\}_{i=1}^{N_s}\) and weights \(\{w_t^i\}_{i=1}^{N_s}\) for \(t = 1, ..., N\) by implementing the particle filter.
    \item Initialize the smoothed weights at \(t = N\) as
    \begin{equation}
        w_{N|N}^i = w_N^i, \quad i = 1, \ldots, M
    \end{equation}
    \item Recursively compute the smoothed weights using \eqref{eq:ps}.
\end{enumerate}

The expected terms in \eqref{eq:Is} can be approximated from the particle smoother outputs as \cite{schon_automatica_2011}:

\begin{subequations} \label{eq:Is_ps}
    \begin{equation} \label{eq:I1_ps}
        I_1 \approx \sum_{i=1}^{N_s} w_{1|N}^{i} \log p_{\theta}(x_{1}^{i}),
    \end{equation}
    \begin{equation} \label{eq:I2_ps}
        I_2 \approx \sum_{t=1}^{N-1} \sum_{i=1}^{N_s} \sum_{j=1}^{N_s} w_{t|N}^{ij} \log p_{\theta}(x_{t+1}^{j} | x_{t}^{i}),
    \end{equation}
    \begin{equation} \label{eq:I3_ps}
        I_3 \approx \sum_{t=1}^{N} \sum_{i=1}^{N_s} w_{t|N}^{i} \log p_{\theta}(y_{1:t} | x_{t}^{i}).
    \end{equation}
\end{subequations}

The weights in \eqref{eq:I2_ps} are determined as follows:

\begin{equation}
    w_{t|N}^{ij} = \frac{w_{t}^{i} w_{t+1|N}^{j} p_{\theta}(x_{t+1}^{j} | x_{t}^{i})}{\sum_{l=1}^{N_s} w_{t}^{l} p_{\theta}(x_{t+1}^{j} | x_{t}^{l})}
\end{equation}

The equations in \eqref{eq:Is_ps} are subsequently utilized to compute the Q function during the E-step of the EM algorithm. Our code was developed by augmenting scripts available in \cite{nordh_jss_2017}.

\begin{table}[ht]
\centering
\caption{Algorithm settings for NLLS, KEM, and PSEM. For all methods, initial parameter estimates for ($F, k_3, k_4$) were drawn from uniform distributions within the ranges listed in Table~\ref{tab:sim_params}, and parameter bounds were set to those same ranges. $F_p$ and $v$ were fixed to their population mean values.}
\label{tab:estimation_settings}
\begin{tabular}{|c|c|c|c|c|c|c|}
\hline
Method & Max Iterations & Estimations & Particles & Trajectories & $p_0$ & $(q, r)$ \\
\hline
NLLS   & 10  & 1 & --  & -- & -- & -- \\
KEM    & 15  & 1 & --  & -- & 10 & $(10,\;0.001)$ \\
PSEM   & 20  & 1 & 150 & 6  & 10 & $(10,\;0.001)$ \\
\hline
\end{tabular}
\end{table}

For the analytical methods (NLLS, KEM, PSEM), although five parameters ($F, k_3, k_4, F_p, v$) were varied in the simulations, only $F, k_3,$ and $k_4$ were estimated, while $F_p$ and $v$ were fixed to their population mean values, consistent with prior literature \cite{herrero_cr_1992, huang_ajphcp_1989}. Estimating all five parameters led to frequent convergence failures due to ill-conditioning of the inverse problem. Accordingly, this setup represents a constrained inverse problem in which the analytical estimators operate under controlled nuisance-parameter misspecification. The estimation settings for each analytical method are summarized in Table~\ref{tab:estimation_settings}. Because NLLS was intended to serve as a baseline method, it was implemented in a standard, out-of-the-box manner with a limited number of iterations and a single estimation.

\subsubsection{Convolutional Neural Network} \label{sec:cnn_methods}
The CNN was trained with supervised learning using mean absolute error (MAE) as the primary loss; mean absolute percentage error (MAPE) was monitored to align with relative error metrics. Training used Adam with a fixed learning rate and mini-batch gradient descent. Convergence was assessed on a validation set; losses stabilized by 10--15 epochs with no divergence (Appendix~\ref{sec:cnn_training}). We used early stopping based on validation MAE or trained to convergence, selecting the best epoch by validation MAE.

\begin{figure}[h]
    \centering
    \includegraphics[width=\columnwidth]{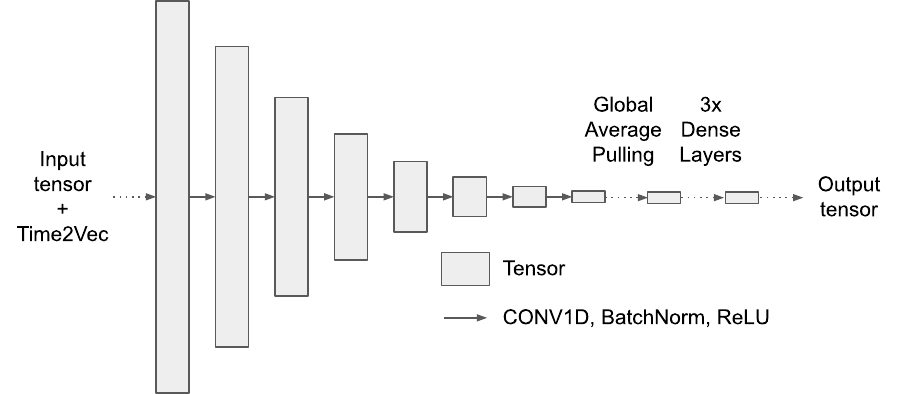}
    \caption{The CNN model consists of seven 1-D convolutional layers, followed by a global average pooling layer, and then three fully connected layers.} \label{fig:cnn}
\end{figure}

The architecture (Fig.~\ref{fig:cnn}) takes a $1{,}024 \times 2$ tensor of interpolated tissue and input TACs normalized by the global maximum. A Time2Vec layer (kernel size 2) encodes temporal information~\cite{kazemi_arxiv_2019}. Seven 1-D convolutional layers (filters 4--256, kernel size 10, padding 4) reduce temporal resolution, followed by global average pooling and three dense layers (16 units each) to estimate \(F\), \(k_3\), and \(k_4\). The output layer uses a sigmoid activation function, which constrains each predicted parameter to the interval $(0,1)$. This is appropriate in our setting, as the simulated parameter ranges for $F$, $k_3$, and $k_4$ (Table \ref{tab:sim_params}) were defined within this interval. No additional rescaling is applied; therefore, the network outputs correspond directly to the physical parameter values, and absolute errors are reported in these same units. The model has 444{,}235 parameters (443{,}155 trainable). Hyperparameter tuning details are in Table~\ref{tab:cnn_params}.

\begin{table}[h]
\caption{The summary of the CNN configuration}
    \centering
    \begin{tabular}{|c|c|}
        \hline
        \textbf{Parameter} & \textbf{Values} \\
        \hline
        Input tensor shape & 1,024 $\times$ 2 \\
        Time2Vec kernel size & 2 \\
        Number of ConvLayers & 7 \\
        Filter size for each ConvLayer & 4, 8, 16, 32, 64, 128, and 256 \\
        ConvLayer padding & 4 \\
        Number of dense layers & 3 \\
        Number of units per each dense layer & 16 \\
        Activation function (hidden layers) & ReLU \\
        Activation function (output layer) & Sigmoid \\
        Loss function & Mean absolute percentage error \\
        Optimizer & Adam \\
        Learning rate & 0.0011 \\
        Batch size & 64 \\
        \hline
    \end{tabular}
    \label{tab:cnn_params}
\end{table}

\subsection{Evaluation methods}
\paragraph{Metrics}
The parameter $(F, k_3, k_4)$ was estimated using the four methods: NLLS, KEM, PSEM, and CNN. The estimations obtained from these methods were then compared to the ground truth values utilized in our simulations. To analyze the accuracy of the estimations, both absolute error and relative error between the estimated and ground truth values were calculated.

\begin{equation} \label{eq:absolute_error}
\text{Absolute error} = \lvert \text{estimated parameter value} - \text{ground truth} \rvert
\end{equation}

\begin{equation} \label{eq:relative_error}
\text{Relative error} = \frac{\lvert \text{estimated parameter value} - \text{ground truth} \rvert}{\text{ground truth}}
\end{equation}

\paragraph{Stratification for near-zero ground truth.}
To explicitly handle the near-zero regime, for each parameter we stratify test cases by the ground-truth magnitude using the 25th percentile threshold computed over the test set. We report results separately for the ``small'' regime ($\theta \leq Q_{25}$) and the ``large'' regime ($\theta > Q_{25}$). The thresholds were $F=0.0169$, $k_3=0.0193$, and $k_4=0.0046$ (units as defined in Table~\ref{tab:sim_params}), yielding $n=50$ cases in the small regime and $n=150$ cases in the large regime for each parameter.

\paragraph{Statistical Significance Analysis}
\label{sec:statistical_analysis}

Performance was compared across methods (NLLS, KEM, PSEM, CNN) using paired simulations and absolute error. For each parameter and noise level (0.8$\times$, 1.0$\times$, 1.2$\times$), we used the Friedman test (simulations as blocks; $n=200$, $k=4$) with Kendall’s $W$ effect size. When significant ($p<0.05$), Wilcoxon signed-rank tests were run with Holm correction across six pairwise comparisons. We report Holm-adjusted $p$-values, median paired differences in absolute error, and rank-biserial correlation. All tests were two-sided ($\alpha=0.05$).

\section{Results}
\subsection{Simulation results}
\begin{figure}[h]
    \centering
    \begin{subfigure}[t]{\linewidth}
        \centering
        \includegraphics[width=\linewidth]{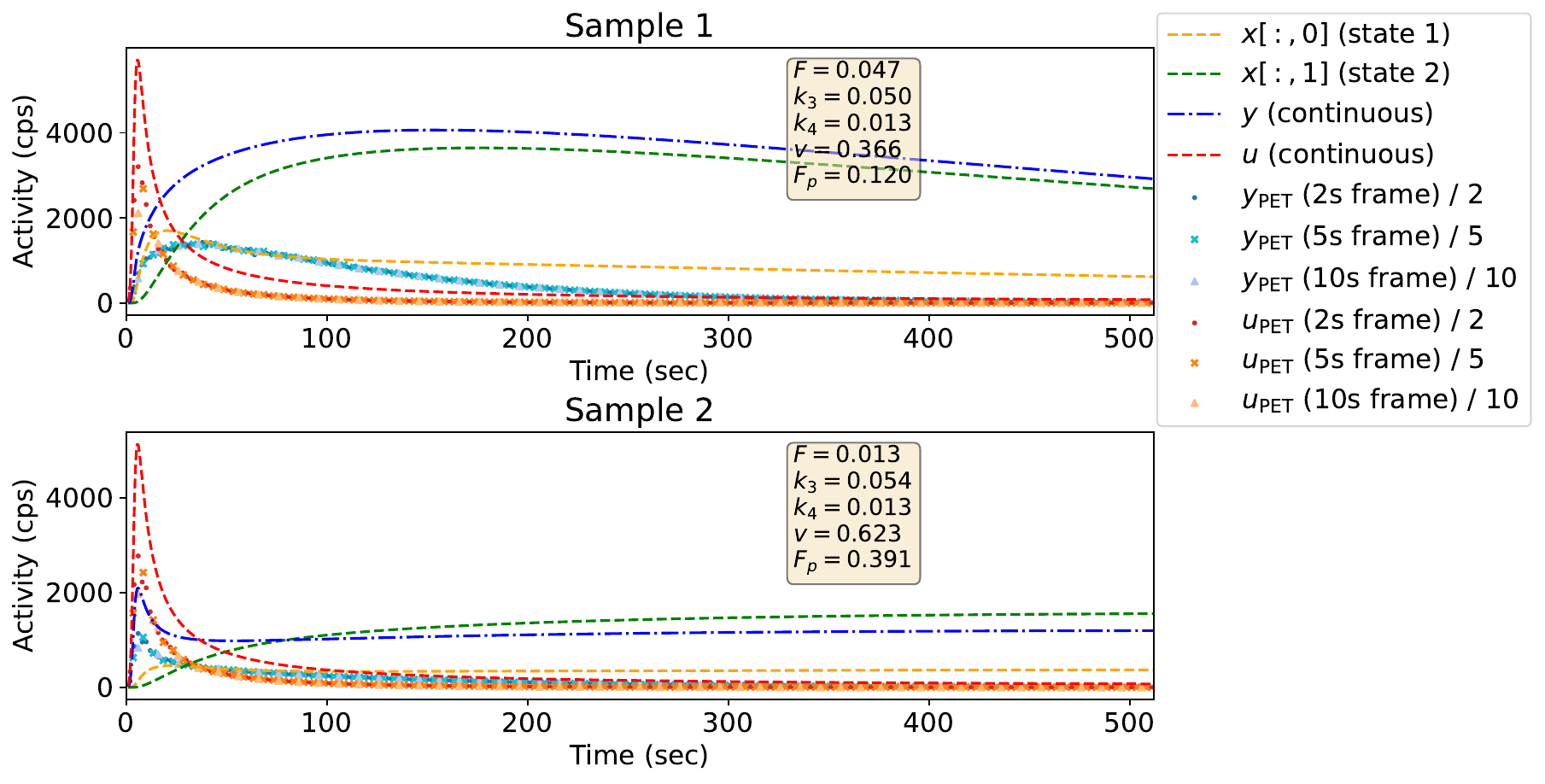}
        \caption*{(a)}
        \phantomcaption
        \label{fig:sim_interp_sim}
    \end{subfigure}

    \vspace{0.5em}

    \begin{subfigure}[t]{\linewidth}
        \centering
        \includegraphics[width=\linewidth]{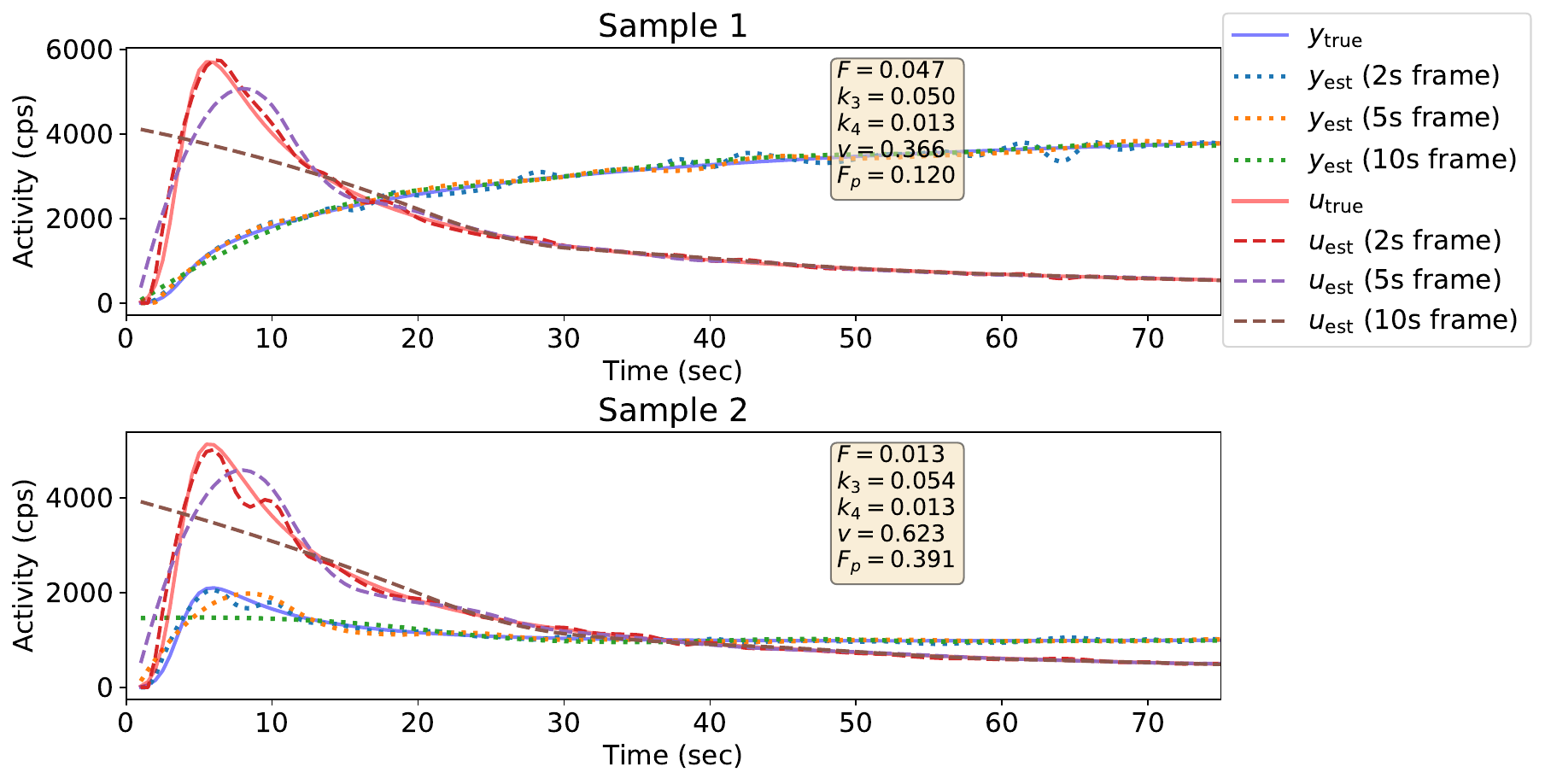}
        \caption*{(b)}
        \phantomcaption
        \label{fig:sim_interp_interp}
    \end{subfigure}

    \caption{Simulation and interpolation results.
    (a) Simulated latent-state and measurement curves for the $^{82}$Rb model; observable activity was computed using \eqref{eq:y} and \eqref{eq:CD_rb82}, and PET/input measurements using \eqref{eq:measurements}.
    (b) Interpolation results for the first 75 seconds; discrepancies are more apparent for 5- and 10-second frames.}
    \label{fig:sim_interp_combined}
\end{figure}

We simulated 512 second acquisitions with frame durations of 2, 5, and 10 seconds. Fig.~\ref{fig:sim_interp_sim} shows the simulation output. Overall, the estimated observable activity and input function (obtained by decay correction and interpolation of PET and input measurements) agree well with ground truth across the full acquisition, albeit with noise. The early-time zoom in Fig.~\ref{fig:sim_interp_interp} highlights deviations in the early frames of the scan, especially for 5- and 10-second frame duration protocols. The interpolation step appears to denoise, as performance is similar across 0.8, 1, and 1.2$\times$ noise levels (Appendix \ref{sec:noise_performance}). 

\subsection{Parameter estimation results} \label{sec:est_results}
\paragraph{Algorithm Runtime}
To assess practical feasibility for clinical translation, we compared the wall-clock time required for parameter estimation \emph{per simulation} across all methods. NLLS, KEM, and PSEM were executed as single-CPU jobs on a high-performance computing cluster with 5~GB memory using Intel Xeon E5-2640 v4 CPU (Broadwell, 2.40 GHz). Runtime was measured as end-to-end wall-clock time and normalized by the number of simulations processed.
Average wall-clock time per simulation was approximately 2.0~s for NLLS, 118~s for KEM, and 468~s for PSEM. 
CNN training and inference was performed on high-performance computing cluster using an NVIDIA A30 GPU in MIG configuration (1g.6gb; 6 GB VRAM per instance) with 3 CPU cores and 16 GB system memory on an AMD EPYC 7543P node (2.75 GHz, 256 GB RAM). CNN training converged within 20-30 minutes. Once trained, CNN inference required approximately 110~s for 200 simulation test set, corresponding to 0.55~s per simulation. All methods were implemented using standard, out-of-the-box configurations without additional hardware-specific optimization. In particular, GPU acceleration was not implemented for the analytical methods, whereas the CNN was implemented in Keras \cite{chollet2015keras}, which leverages GPU acceleration. Therefore, the reported runtimes may reflect differences in implementation and hardware utilization in addition to algorithmic complexity. 

\begin{figure}
    \centering
    \includegraphics[width=\linewidth]{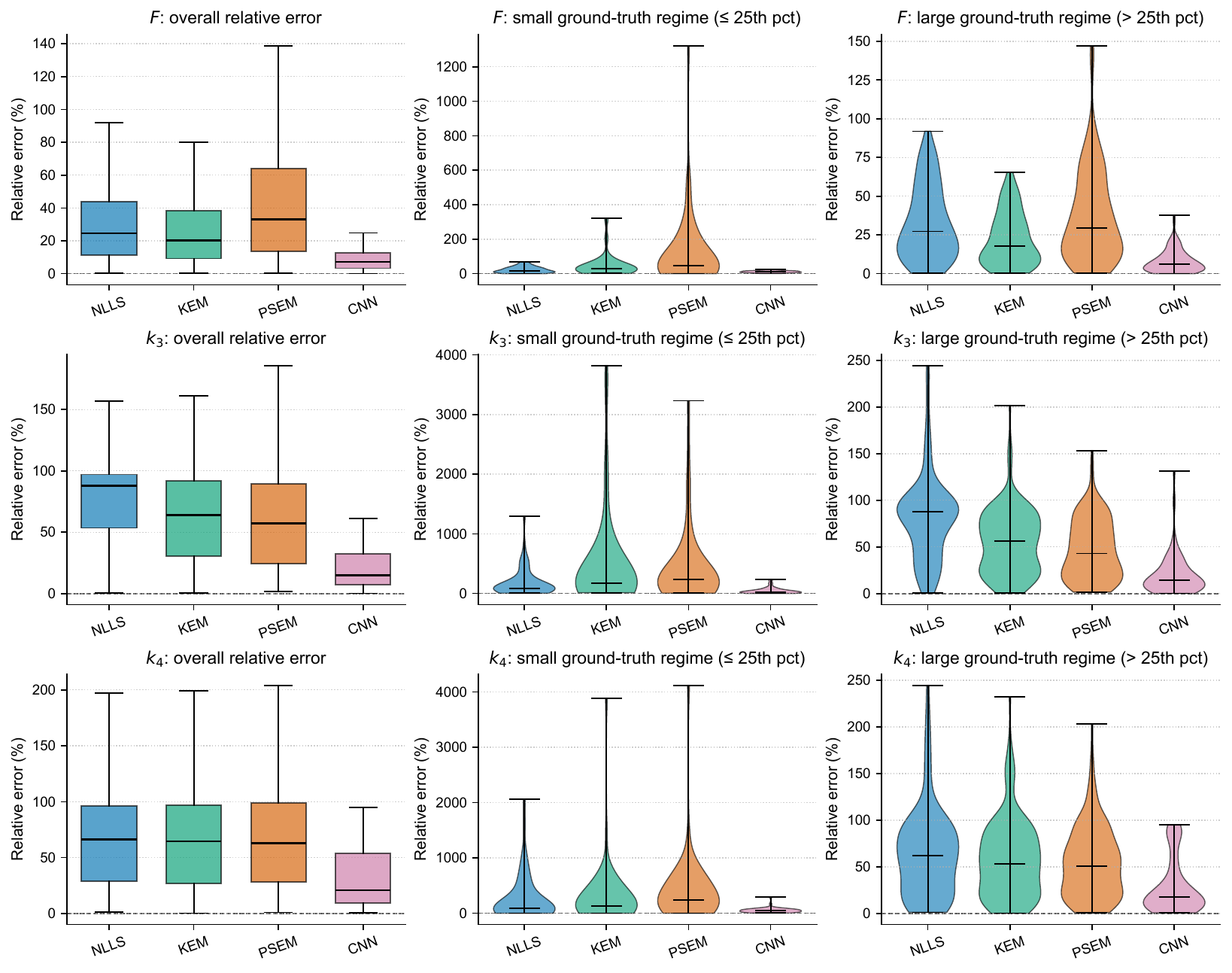}
    \caption{Relative error distributions at 1.0$\times$ noise and 2~s frame duration.
    Left: overall relative errors across all simulations shown as boxplots; the central line indicates the median, the box spans the interquartile range (25th–75th percentiles), whiskers extend to 1.5$\times$ the interquartile range, and outliers are omitted.
    Center: relative error distributions for the small ground-truth regime ($\leq$25th percentile) shown as violin plots; the violin width reflects the kernel density of the data, the central marker denotes the median, and vertical lines indicate the full data range (minimum to maximum).
    Right: relative error distributions for the large ground-truth regime ($>$25th percentile) shown as violin plots with the same conventions.}
    \label{fig:error_combined}
\end{figure}

\paragraph{Estimation Errors}
Table \ref{tab:abs-stacked} summarizes absolute and relative errors for {the parameter settings in Table \ref{tab:estimation_settings}} and corresponding box plots in Fig. \ref{fig:error_combined}. The CNN showed the best result among the other methods in terms of both percentage and absolute error for all time frame durations. 

\vspace{0.5em}
\begin{table}[!htbp]
\caption{Absolute and relative errors across time frames (mean $\pm$ SEM, 200 test samples). Absolute-error units for $F$, $k_3$, and $k_4$ were ml/s, 1/s, 1/s, respectively.}
    \centering
    \setlength{\tabcolsep}{4pt}
    \renewcommand{\arraystretch}{1.1}
    \begin{tabular}{|c|c|c|c|c|}
        \hline
        \multicolumn{5}{|c|}{\textbf{Absolute error}} \\ \hline
        Method & Parameter & 2 s & 5 s & 10 s \\ \hline
        NLLS & $F$  & 0.0105 $\pm$ 0.0008 & 0.0105 $\pm$ 0.0008 & 0.0105 $\pm$ 0.0008 \\
         & $k_3$  & 0.0282 $\pm$ 0.0012 & 0.0283 $\pm$ 0.0012 & 0.0281 $\pm$ 0.0012 \\
         & $k_4$  & 0.0058 $\pm$ 0.0003 & 0.0058 $\pm$ 0.0003 & 0.0058 $\pm$ 0.0003 \\
        \hline
        KEM & $F$  & 0.0083 $\pm$ 0.0006 & 0.0089 $\pm$ 0.0006 & 0.0081 $\pm$ 0.0006 \\
         & $k_3$  & 0.0241 $\pm$ 0.0012 & 0.0242 $\pm$ 0.0012 & 0.0241 $\pm$ 0.0012 \\
         & $k_4$  & 0.0057 $\pm$ 0.0003 & 0.0057 $\pm$ 0.0003 & 0.0057 $\pm$ 0.0003 \\
        \hline
        PSEM & $F$  & 0.0138 $\pm$ 0.0010 & 0.0134 $\pm$ 0.0009 & 0.0140 $\pm$ 0.0010 \\
         & $k_3$  & 0.0215 $\pm$ 0.0011 & 0.0212 $\pm$ 0.0011 & 0.0213 $\pm$ 0.0011 \\
         & $k_4$  & 0.0059 $\pm$ 0.0003 & 0.0058 $\pm$ 0.0003 & 0.0057 $\pm$ 0.0003 \\
        \hline
        CNN & $F$  & 0.0027 $\pm$ 0.0002 & 0.0016 $\pm$ 0.0001 & 0.0012 $\pm$ 0.0001 \\
         & $k_3$  & 0.0065 $\pm$ 0.0004 & 0.0071 $\pm$ 0.0006 & 0.0087 $\pm$ 0.0007 \\
         & $k_4$  & 0.0026 $\pm$ 0.0002 & 0.0020 $\pm$ 0.0002 & 0.0015 $\pm$ 0.0001 \\
        \hline
        \multicolumn{5}{|c|}{\textbf{Relative error (\%)}} \\ \hline
        Method & Parameter & 2 s & 5 s & 10 s \\ \hline
        NLLS & $F$  & 30.15 $\pm$ 1.62 & 30.18 $\pm$ 1.61 & 30.31 $\pm$ 1.60 \\
         & $k_3$  & 107.27 $\pm$ 9.41 & 107.37 $\pm$ 9.50 & 107.04 $\pm$ 9.57 \\
         & $k_4$  & 118.32 $\pm$ 16.39 & 118.25 $\pm$ 16.37 & 117.43 $\pm$ 16.25 \\
        \hline
        KEM & $F$  & 29.19 $\pm$ 2.75 & 29.57 $\pm$ 2.46 & 30.37 $\pm$ 2.95 \\
         & $k_3$  & 166.00 $\pm$ 31.19 & 166.25 $\pm$ 31.09 & 166.09 $\pm$ 31.21 \\
         & $k_4$  & 121.37 $\pm$ 21.48 & 122.27 $\pm$ 21.48 & 120.33 $\pm$ 21.43 \\
        \hline
        PSEM & $F$  & 58.99 $\pm$ 8.17 & 55.52 $\pm$ 7.10 & 60.77 $\pm$ 8.51 \\
         & $k_3$  & 146.02 $\pm$ 25.91 & 147.40 $\pm$ 26.28 & 148.55 $\pm$ 26.64 \\
         & $k_4$  & 144.06 $\pm$ 23.90 & 141.14 $\pm$ 23.55 & 133.06 $\pm$ 22.01 \\
        \hline
        CNN & $F$  & 8.78 $\pm$ 0.49 & 7.05 $\pm$ 0.54 & 4.98 $\pm$ 0.31 \\
         & $k_3$  & 26.05 $\pm$ 2.44 & 23.47 $\pm$ 2.20 & 25.50 $\pm$ 1.86 \\
         & $k_4$  & 34.34 $\pm$ 2.52 & 22.54 $\pm$ 1.54 & 22.76 $\pm$ 2.04 \\
        \hline
    \end{tabular}
    \label{tab:abs-stacked}
\end{table}

\paragraph{Stratification for near-zero ground truth.}
Fig.~\ref{fig:error_combined} summarizes overall and stratified error distributions. Analytical methods (NLLS, KEM, PSEM) exhibit highly dispersed and heavy-tailed relative error distributions in the near-zero regime, reflecting sensitivity to small absolute deviations. In contrast, the CNN maintains comparatively stable error distributions across both regimes, indicating improved robustness when estimating parameters with low true values.

\subsubsection{Paired Method Comparison Results}
\label{sec:statistical_results}
Friedman tests were significant in all conditions, leading to Wilcoxon signed-rank tests on all parameters. Kendall’s $W$ ranged from 0.146 to 0.432, indicating small to moderate agreement among methods ($W \geq 0.1$: small, $W \geq 0.3$: medium, $W \geq 0.5$: large \cite{fiel_peres_effect_nodate}). The results for 2~s frame time and default noise level are shown in Table~\ref{tab:posthoc_1x_2s_sorted}, and these comparisons generalize across different frame durations and noise levels.

\paragraph{Blood flow ($F$).}
For all noise levels, the CNN achieved significantly lower absolute error than NLLS, KEM, and PSEM (Holm-adjusted $p \ll 10^{-23}$). Additional but smaller differences were observed among the conventional methods, with KEM outperforming NLLS and PSEM.

\paragraph{Parameter $k_3$.}
Across all noise levels, CNN again demonstrated significantly lower absolute error than NLLS, KEM, and PSEM (Holm-adjusted $p \ll 10^{-23}$). Among the conventional approaches, PSEM showed modest but statistically significant improvements over NLLS and KEM, while differences between NLLS and KEM were smaller but still statistically significant.

\paragraph{Parameter $k_4$.}
CNN achieved significantly lower absolute error than NLLS, KEM, and PSEM at all noise levels (Holm-adjusted $p \ll 10^{-15}$). In contrast, no statistically significant differences were observed among NLLS, KEM, and PSEM after Holm correction, indicating comparable performance among the analytical methods for this parameter.

\begin{table}[t]
\centering
\caption{Significant pairwise Wilcoxon signed-rank comparisons (Holm-corrected) at 1.0$\times$ noise and 2\,s frame time. 
Values show $\mathrm{median}(|e_A|-|e_B|)$; positive values indicate lower error for method $B$ (winner in \textbf{bold}).}
\label{tab:posthoc_1x_2s_sorted}
\setlength{\tabcolsep}{5pt}
\renewcommand{\arraystretch}{1.15}

\begin{tabular}{lccc}
\hline
Parameter & Comparison $(A\ \mathrm{vs.}\ B)$ & Median $\Delta|e|$ & $p_{\mathrm{Holm}}$ \\
\hline

$F$ 
& PSEM vs \textbf{CNN} & $+6.24\times10^{-3}$ & $5.8\times10^{-28}$ \\
& NLLS vs \textbf{CNN} & $+4.61\times10^{-3}$ & $3.5\times10^{-27}$ \\
& KEM vs \textbf{CNN}  & $+2.43\times10^{-3}$ & $7.8\times10^{-24}$ \\
& \textbf{KEM} vs PSEM & $-2.74\times10^{-3}$ & $1.7\times10^{-9}$ \\
& NLLS vs \textbf{KEM} & $+1.45\times10^{-3}$ & $5.8\times10^{-8}$ \\
[4pt]

$k_3$
& NLLS vs \textbf{CNN} & $+1.89\times10^{-2}$ & $4.7\times10^{-32}$ \\
& KEM vs \textbf{CNN}  & $+1.45\times10^{-2}$ & $6.8\times10^{-27}$ \\
& PSEM vs \textbf{CNN} & $+1.11\times10^{-2}$ & $5.4\times10^{-24}$ \\
& NLLS vs \textbf{PSEM} & $+8.26\times10^{-3}$ & $2.0\times10^{-5}$ \\
& KEM vs \textbf{PSEM}  & $+1.42\times10^{-3}$ & $2.5\times10^{-5}$ \\
& NLLS vs \textbf{KEM}   & $+2.38\times10^{-3}$ & $2.1\times10^{-4}$ \\
[4pt]

$k_4$
& PSEM vs \textbf{CNN} & $+2.80\times10^{-3}$ & $7.5\times10^{-18}$ \\
& KEM vs \textbf{CNN}  & $+2.57\times10^{-3}$ & $2.3\times10^{-17}$ \\
& NLLS vs \textbf{CNN} & $+2.30\times10^{-3}$ & $1.2\times10^{-16}$ \\
\hline
\end{tabular}
\end{table}

\subsection{Out of Distribution Results}
\begin{figure}
    \centering
    \includegraphics[width=\linewidth]{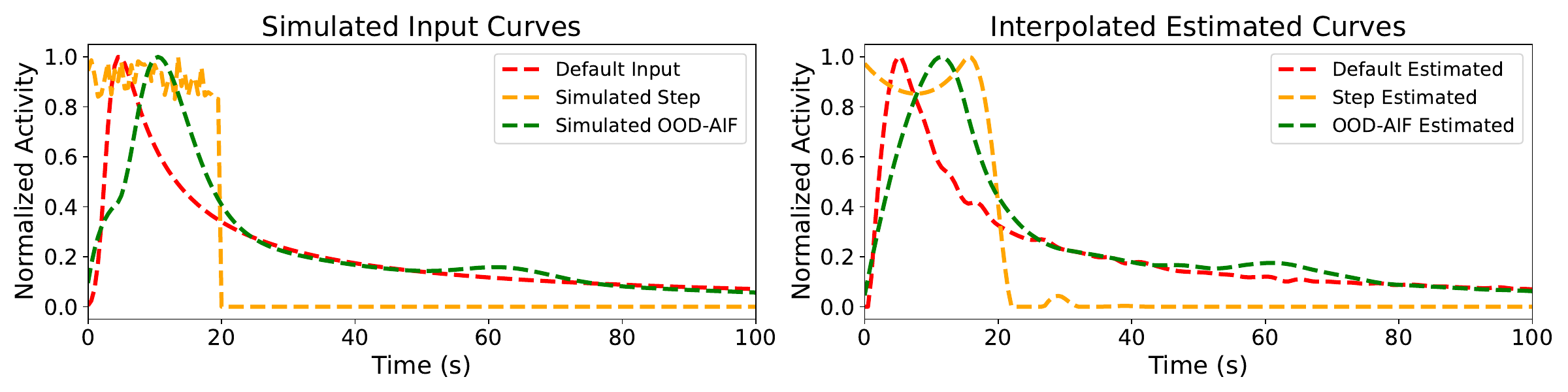}

    \vspace{0.5em}

    \includegraphics[width=\linewidth]{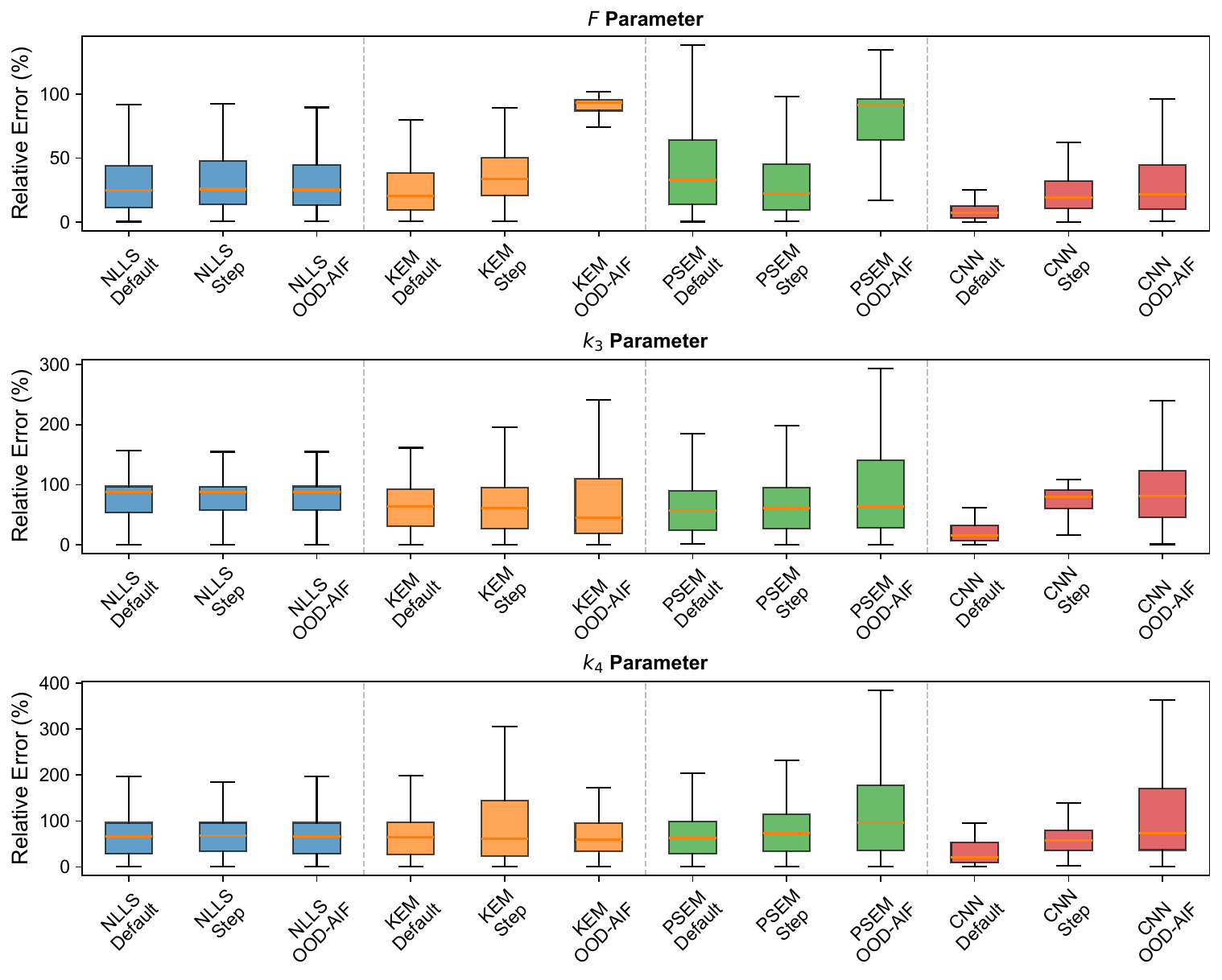}

    \caption{OOD input-function analysis.
    Top: default, step, and OOD-AIF input-function simulation and estimation.
    Bottom: box plots of relative error for $F$, $k_3$, and $k_4$ at 5~s frames and 1.0$\times$ noise across methods. 
    In each box, the central line denotes the median, the box bounds indicate the interquartile range (25th–75th percentiles), and the whiskers extend to 1.5$\times$ the interquartile range; outliers are not shown.}
    \label{fig:ood_combined}
\end{figure}

 When we changed the input function to step and OOD-AIF, the resulting simulation and estimation as well as relative errors across the different input functions compared to the default input function are shown in Fig. \ref{fig:ood_combined} and Table \ref{tab:ood_stats}.

\begin{table}[t] 
\centering
\caption{Mean relative error (\%) under step and OOD-AIF input conditions  (mean $\pm$ SEM, 200 test samples), with the change relative to default input shown in parentheses (percentage-point difference).}
\label{tab:ood_stats}
\setlength{\tabcolsep}{4pt}
\renewcommand{\arraystretch}{1.2}
\begin{tabular}{llcccc}
\hline
Condition & Param & NLLS & KEM & PSEM & CNN \\
\hline
\multirow{3}{*}{Step}
 & $F$ &
 \begin{tabular}{@{}c@{}}31.55$\pm$1.59 \\ (+1.37)\end{tabular} &
 \begin{tabular}{@{}c@{}}43.51$\pm$3.88 \\ (+13.94)\end{tabular} &
 \begin{tabular}{@{}c@{}}48.65$\pm$6.17 \\ (+7.06)\end{tabular} &
 \begin{tabular}{@{}c@{}}24.58$\pm$1.40 \\ (+17.53)\end{tabular} \\

 & $k_3$ &
 \begin{tabular}{@{}c@{}}106.77$\pm$9.31 \\ (-0.60)\end{tabular} &
 \begin{tabular}{@{}c@{}}154.89$\pm$28.66 \\ (-11.36)\end{tabular} &
 \begin{tabular}{@{}c@{}}174.66$\pm$32.71 \\ (-90.09)\end{tabular} &
 \begin{tabular}{@{}c@{}}75.08$\pm$2.27 \\ (+51.61)\end{tabular} \\

 & $k_4$ &
 \begin{tabular}{@{}c@{}}110.74$\pm$15.80 \\ (-7.51)\end{tabular} &
 \begin{tabular}{@{}c@{}}173.53$\pm$27.11 \\ (+51.26)\end{tabular} &
 \begin{tabular}{@{}c@{}}177.78$\pm$31.94 \\ (+30.84)\end{tabular} &
 \begin{tabular}{@{}c@{}}68.69$\pm$4.11 \\ (+46.15)\end{tabular} \\
\hline
\multirow{3}{*}{OOD-AIF}
 & $F$ &
 \begin{tabular}{@{}c@{}}30.67$\pm$1.59 \\ (+0.49)\end{tabular} &
 \begin{tabular}{@{}c@{}}89.88$\pm$2.39 \\ (+60.31)\end{tabular} &
 \begin{tabular}{@{}c@{}}182.22$\pm$32.99 \\ (+140.63)\end{tabular} &
 \begin{tabular}{@{}c@{}}31.32$\pm$2.06 \\ (+24.27)\end{tabular} \\

 & $k_3$ &
 \begin{tabular}{@{}c@{}}107.74$\pm$9.22 \\ (+0.37)\end{tabular} &
 \begin{tabular}{@{}c@{}}160.00$\pm$26.27 \\ (-6.25)\end{tabular} &
 \begin{tabular}{@{}c@{}}208.44$\pm$34.95 \\ (-56.31)\end{tabular} &
 \begin{tabular}{@{}c@{}}201.40$\pm$34.85 \\ (+177.93)\end{tabular} \\

 & $k_4$ &
 \begin{tabular}{@{}c@{}}115.18$\pm$16.10 \\ (-3.07)\end{tabular} &
 \begin{tabular}{@{}c@{}}148.26$\pm$23.01 \\ (+25.99)\end{tabular} &
 \begin{tabular}{@{}c@{}}224.43$\pm$35.53 \\ (+77.49)\end{tabular} &
 \begin{tabular}{@{}c@{}}152.93$\pm$16.48 \\ (+130.39)\end{tabular} \\
\hline
\end{tabular}
\end{table}

\section{Discussion}
\paragraph{Simulation design and identifiability} Fig. \ref{fig:sim_interp_sim} illustrates how the kinetic parameters affects the simulated time-activity curves. First, we can notice that the activity of State 1 (fast exchangeable state) accumulates more when $v$ is higher and $k_3$ is lower. Similarly, high $k_3$ values and low $k_4$ values result in high activity in State 2 (slow exchangeable state). PET measurements and input {function} measurements decay over time, and we added noise estimated from real datasets to these measurements, resulting in signal degradations. Lastly, the changes in signal are reduced over time, which indicates that most of the extractable information regarding kinetics is stored in the early stage measurements.

This is supported by a local conditioning analysis based on the Jacobian and Fisher information matrix, which shows that estimation of $(F,k_3,k_4)$ is intrinsically ill-conditioned in low-exchange regimes, with strong parameter correlations and large variance amplification (see Appendix \ref{sec:cond_appendix}). Accordingly, the large errors observed for analytical estimators in these regimes reflect fundamental limitations of the inverse problem rather than deficiencies of the optimization procedure as seen in Fig. \ref{fig:error_combined}. 

\paragraph{Recovery of observable signals }Fig.~\ref{fig:sim_interp_interp} show the estimated observable activities and input functions recovered from the PET and input measurements. Overall, the estimated curves agree well with the ground truths, with noise decreasing over time as signal variations become less abrupt. However, the estimated $y_{\text{est}}$ and $u_{\text{est}}$ deviate from the ground-truth $y_{\text{true}}$ and $u_{\text{true}}$ during the first 15 seconds (Fig.~\ref{fig:sim_interp_interp}), although this has little impact on the overall metrics, which remain consistent across frame durations seen in Sec. \ref{sec:noise_performance}. The interpolation performance is a bit weaker with the out of distribution test functions shown at the top figure of Fig. \ref{fig:ood_combined}.

\paragraph{Estimation performance in distribution}Table \ref{tab:abs-stacked} and Fig. \ref{fig:error_combined} show the estimation performances of each proposed method in absolute and relative errors. The absolute and relative errors results show that the CNN performed better than the other three approaches for all three kinetic parameters by having the lowest estimation errors for all three parameters. The absolute errors of KEM for parameter $F$ were less than that of NLLS, whereas those of PSEM underperformed compared to NLLS. This may indicate that the PSEM needs hand-tuning to be employed in practice. KEM and PSEM performed better than NLLS in estimating $k_3$, but the differences among the results of three methods estimating $k_4$ had no statistical significance. A snippet of the post-hoc statistical analysis is shown in Table \ref{tab:posthoc_1x_2s_sorted}, confirming the CNN’s outperformance across all three parameters.  Among the analytical approaches, for $F$, KEM outperformed PSEM and NLLS, with no statistically significant difference between NLLS and PSEM. For $k_3$, PSEM outperformed both NLLS and KEM (with KEM > NLLS), whereas for $k_4$ there was no statistically significant difference among the analytical methods. These conclusions were consistent across all frame durations and noise levels.

\paragraph{CNN robustness to nuisance-parameter misspecification.}
The superior performance of the CNN under the evaluated conditions can be interpreted in the context of nuisance-parameter uncertainty. The plasma fraction $F_p$ and volume fraction $v$ were treated as nuisance parameters and were not estimated, but instead fixed to constant values for the model-based estimators, introducing controlled model misspecification relative to the simulation model. Because these parameters enter explicitly into the forward model, misspecification can bias likelihood-based estimation and degrade performance. In contrast, the CNN does not rely on explicit values of $F_p$ or $v$ at inference and instead learns a direct mapping from measured time--activity curves to the kinetic parameters $(F, k_3, k_4)$. As a result, the CNN is less sensitive to variability in nuisance parameters that is present in the data but not explicitly modeled during estimation. Accordingly, this comparison evaluates robustness to model misspecification rather than purely intrinsic estimation accuracy. This distinction is supported by the results in Appendix~\ref{sec:true_other_params}, where providing the true $F_p$ and $v$ substantially improves analytical method performance and allows NLLS to outperform the CNN.

\paragraph{PSEM performance}
While the Rb-82 two-tissue compartment model is linear, we present PSEM here as a proof-of-concept baseline within a controlled setting, allowing direct comparison with KEM under identical assumptions. The mixed performance observed highlights that increased model flexibility does not necessarily translate to improved accuracy in predominantly linear problems, an insight we believe is valuable to report.
{Fig. }\ref{fig:est_vs_iter} located in Appendix \ref{sec:appendix_psem}, depicts the convergence of the ten PSEM estimations towards the ground truth value of parameter $F$. All estimations converge monotonically to the ground truth value, demonstrating an indirect manifestation of Jensen's inequality within the EM algorithm. However, it is important to note that while this sample highlights the algorithm's effective performance in certain instances, the accuracy of PSEM outputs deteriorates when estimating multiple parameters and also varies significantly based on ground-truth values, as reported in Section \ref{sec:est_results}. The improved $k_3$ estimation and degraded $F$ estimation observed with PSEM likely reflect parameter-dependent identifiability, as particle smoothing can enhance robustness for weakly observable exchange parameters while introducing sampling variability that can slightly degrade accuracy for strongly constrained parameters such as $F$.

\paragraph{Generalization}%
While the CNN demonstrates strong performance under in-distribution conditions, its performance degrades substantially under out-of-distribution (OOD) settings, as shown in Fig.~\ref{fig:ood_combined} and Table~\ref{tab:ood_stats}. In these cases, the CNN may perform worse than analytical methods. This behavior reflects the fact that the CNN learns a data-driven mapping tied to the training distribution, whereas analytical methods incorporate mechanistic model constraints that provide greater robustness to changes in acquisition conditions and input functions.

For the CNN, generalization to a new tracer, different framing protocol, or different input function (Fig.~\ref{fig:ood_combined} and Table~\ref{tab:ood_stats}) requires retraining and, if necessary, adjustment of the output layer to match the number of kinetic parameters. While our experiments focused on $^{82}$Rb, the analytical methods (NLLS, KEM, and PSEM) are generally applicable to other tracers as long as the linear dynamical system representation of the tracer and plausible parameter ranges are defined. In this work, the CNN is employed in a manner tailored to dynamic $^{82}$Rb imaging.

Given the observed out-of-distribution (OOD) behavior (Fig.~\ref{fig:ood_combined} and Table~\ref{tab:ood_stats}) and the requirement for retraining across framing schemes, the CNN method is positioned as a starting point for incorporating deep learning into Rb-82 PET kinetic parameter estimation. Importantly, the CNN is trained and evaluated under controlled simulation conditions, and its performance is therefore conditional on similar acquisition settings. In clinical practice, variability in noise levels, input functions, and scanner protocols may lead to domain shift and degraded performance. Furthermore, generalization across sites, scanners, and reconstruction settings has not been established. Addressing the domain-shift limitation of the CNN will require improving robustness to acquisition variability, for example through hybrid approaches that incorporate model-based constraints into the learning framework \cite{pan_kinetic_2024}. In its current form, the need to retrain the CNN for each tracer and framing protocol presents a barrier to clinical adoption. Once trained, the CNN offers faster inference and greater robustness to nuisance-parameter misspecification compared with analytical methods. 

\paragraph{Clinical Impact}%
Across typical flow values within the simulated range, the CNN achieved mean absolute errors corresponding to approximately 4–9\% relative error. This error magnitude is substantially smaller than commonly used myocardial perfusion category separations, which are typically on the order of 20–30\% \cite{murthy_association_2012, gould_anatomic_2013, murthy_clinical_2018}. However, these comparisons are based on controlled simulations that do not incorporate motion, spillover, reconstruction bias, or inter-patient variability, and no formal classification analysis was performed. Therefore, this observation should be interpreted as a theoretical implication of the error scale rather than demonstrated clinical impact. Furthermore, additional metrics more directly tied to clinical interpretation could be considered in future work. These include regional classification accuracy for normal versus hypoperfused myocardium based on established uptake or flow thresholds, test–retest repeatability, and agreement in defect extent or severity relative to reference methods. Evaluating performance under stress conditions and low-count acquisitions would further clarify the potential clinical impact of the proposed approaches, but such analyses are beyond the scope of the present work. 

In this work, NLLS was evaluated using a standard configuration without explicit regularization and with default convergence criteria (Table~\ref{tab:estimation_settings}) and population mean values assumption for nuisance parameters, reflecting a representative baseline implementation. Although results were obtained from a single initialization, a retrospective multi-start analysis using interior points in the feasible parameter space showed that all initializations converged to the same solution (Appendix~\ref{sec:nlls_init}), indicating low sensitivity to initialization under the evaluated conditions. The primary limitation of NLLS in this setting is therefore not optimization difficulty, but sensitivity to model misspecification, particularly due to fixed or imperfectly specified nuisance parameters (Appendix~\ref{sec:true_other_params}). In contrast, the CNN was trained with tuned hyperparameters and implicitly accounts for variability in these parameters through data-driven learning. The reported results should be interpreted as comparisons between a representative implementation of NLLS and a trained CNN, rather than a comparison of fully optimized methods. Finally, this study does not include uncertainty quantification. Developing a consistent and fair framework for uncertainty estimation across analytical and learning-based methods remains an open problem and an important direction for future work.

In this first application of the PSEM approach to kinetic modeling, we used the hidden states of the state-space model as the latent variables in the E-step. Future improvements may be achieved by moving selected kinetic parameters into the latent space and solving the resulting problem with a Rao–Blackwellized PSEM, which has been shown to enhance the accuracy of particle smoothers \cite{lindsten_assp_2013}. An additional advantage of PSEM is its applicability to nonlinear kinetic models \cite{doucet2000SMC,andrieu2010PMCMC,lindsten2013RBPS}. Repartitioning, Rao–Blackwellization, and extending PSEM to nonlinear kinetic models are promising directions for future work. 

\section{Conclusions}
In this work, we investigated, for the first time, particle smoother-based Expectation--Maximization (PSEM) and convolutional neural network (CNN) approaches for kinetic parameter estimation in simulated $^{82}$Rb dynamic PET. Compared with nonlinear least squares (NLLS) and Kalman smoother-based EM (KEM), the CNN consistently achieved the highest accuracy across parameters and frame durations for in-distribution samples. PSEM showed parameter-dependent performance, improving $k_3$ estimation but underperforming for myocardial blood flow $F$. Future work will extend these evaluations to additional tracers, more realistic simulations, and patient data to assess robustness and generalizability. Further methodological development of PSEM and expanded evaluation using clinically interpretable performance measures will help clarify the trade-offs between accuracy, robustness, interpretability, and potential clinical relevance.

\ifblind
\else
\section{Acknowledgments}
Sarah was supported by The Sun-Pan Family Fellowship Fund and is the Mark and Mary Stevens Interdisciplinary Graduate Fellow. The authors would like to thank Dr. Mojtaba Jafaritadi for helpful discussions on deep learning ideas.

\section{Conflict of Interest Statement}
The authors declare no conflict of interest.
\fi

\section{Appendix} 
\subsection{Noise estimation} \label{sec:noise_estimation}

A simplified but realistic PET noise model was derived from $^{82}$Rb cardiac data acquired on a Discovery 690 PET/CT (GE Healthcare). Noise statistics were estimated from a $36\times36$ pixel myocardial ROI normalized to mean 1, and variability was summarized using the variance-to-mean ratio (VMR). Simulations used noise levels corresponding to clinically relevant short frame durations of 2, 5, and 10~s, with empirically measured VMR values of 0.0767, 0.0397, and 0.0280, respectively. For each frame with mean signal $\mu_n$, noise was added to the ROI-averaged measurement with variance scaled by the VMR and the number of ROI pixels $N$. Estimating noise from a single ROI and assuming pixel independence is approximate; the measured variance also reflects spatial heterogeneity and reconstruction correlations. Nevertheless, this empirically derived model provides a reasonable approximation of PET noise for simulation studies and is sufficient for relative method comparison.

\subsection{PSEM estimation along the iterations}
\label{sec:appendix_psem}
\begin{figure}[h]
    \centering
    \includegraphics[width=\columnwidth]{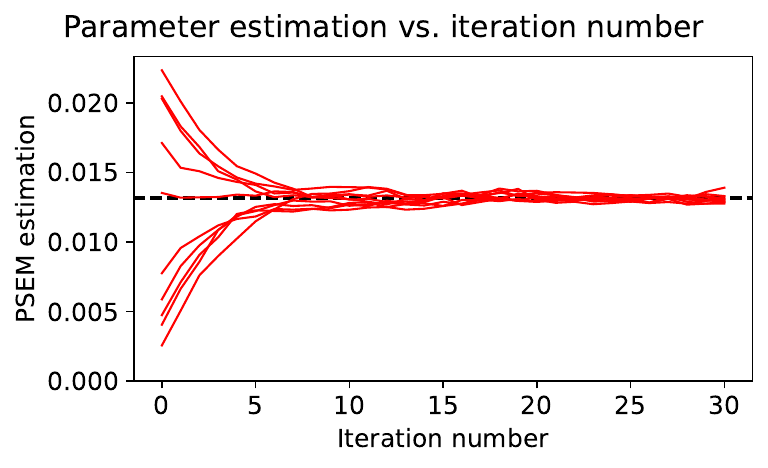}
    \caption{A plot of PSEM estimation across iterations for {parameter $F$} (where all other parameters $k_3, k_4, F_p, v$ were set to their true value) for a sample simulation is shown: In this example, ten independent estimations were performed, each with a random initial guess. {The black dotted line indicates the true value of the estimated parameter.}} 
    \label{fig:est_vs_iter}

\end{figure}

Fig.~\ref{fig:est_vs_iter} shows PSEM convergence for one simulation. Only $F$ is estimated; $k_3, k_4, F_p, v$ are fixed to ground truth. PSEM converges to the true $F$ across random initializations.

\subsection{Conditioning and identifiability analysis}
\label{sec:cond_appendix}
We assessed identifiability of $(F,k_3,k_4)$ using local sensitivities (Jacobian) and the Fisher information matrix (FIM). Sensitivities were computed by integrating an augmented ODE system at representative parameter points. Conditioning was summarized by FIM condition number and induced parameter correlations; large values indicate poor identifiability. Table~\ref{tab:conditioning} shows severe ill-conditioning in low-exchange regimes (condition numbers $>10^{6}$).

\begin{table}[t]
\centering
\caption{Conditioning metrics for representative kinetic parameter regimes using $C=[0.5,0.5]$ and $V_1=5$. Reported values include the condition number of the Fisher information matrix (FIM) and the maximum absolute off-diagonal parameter correlation. Large condition numbers and near-unit correlations indicate severe ill-conditioning and limited practical identifiability.}
\label{tab:conditioning}
\begin{tabular}{lcc}
\hline
Parameter regime & $\mathrm{cond}(\mathrm{FIM})$ & $\max|\rho|$ \\
\hline
Typical (log-mid)          & $2.19\times10^{4}$ & $0.95$ \\
High flow (typical)        & $2.68\times10^{2}$ & $0.93$ \\
Small $k_3$                & $6.36\times10^{4}$ & $0.98$ \\
Small $k_4$                & $8.32\times10^{5}$ & $0.91$ \\
Small $k_3$, small $k_4$   & $3.00\times10^{6}$ & $0.98$ \\
High flow, small $k_4$     & $1.41\times10^{4}$ & $0.87$ \\
All high                   & $6.28\times10^{3}$ & $1.00$ \\
All low                    & $3.44\times10^{8}$ & $0.98$ \\
\hline
\end{tabular}
\end{table}

\subsection{CNN Training Convergence and Validation Performance}
\label{sec:cnn_training}
\begin{figure}
    \centering
    \includegraphics[width=\linewidth]{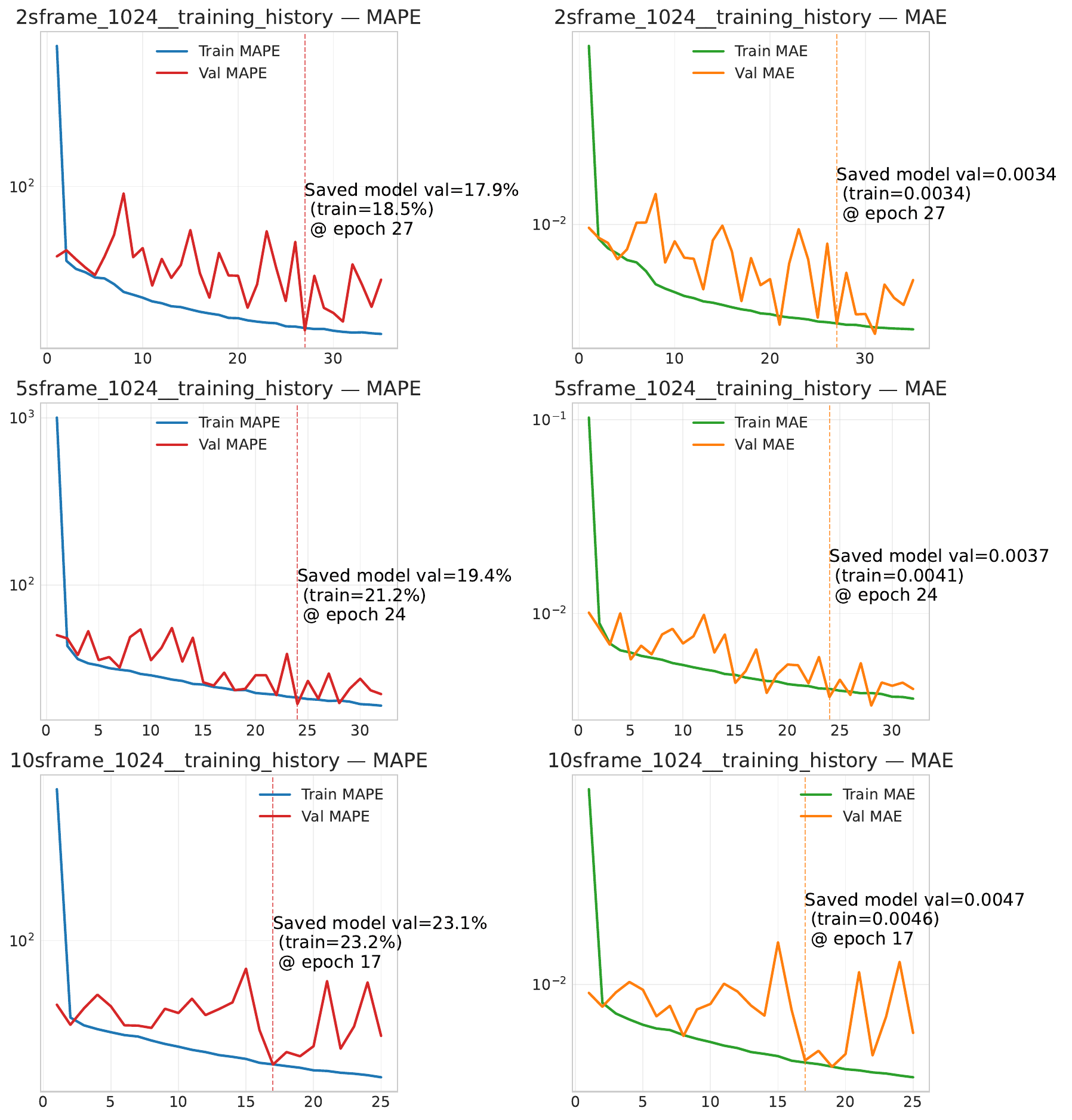}
    \caption{Training and validation loss curves for the CNN model with log scale. (a) Mean absolute error (MAE) and (b) mean absolute percentage error (MAPE) as a function of training epoch. Validation loss converged within approximately 10--15 epochs, indicating stable generalization and no evidence of overfitting.}
    \label{fig:cnn_training_curves}
\end{figure}

Training and validation loss curves (Fig.~\ref{fig:cnn_training_curves}) show rapid early improvement and plateau by 10--15 epochs. Training MAE continued to decrease slightly thereafter, while validation MAE stabilized, indicating convergence without overfitting. The final model was selected at the epoch with minimum validation MAE.

\subsection{Performance over noise levels} \label{sec:noise_performance}
\begin{figure}
    \centering
    \includegraphics[width=0.75\linewidth]{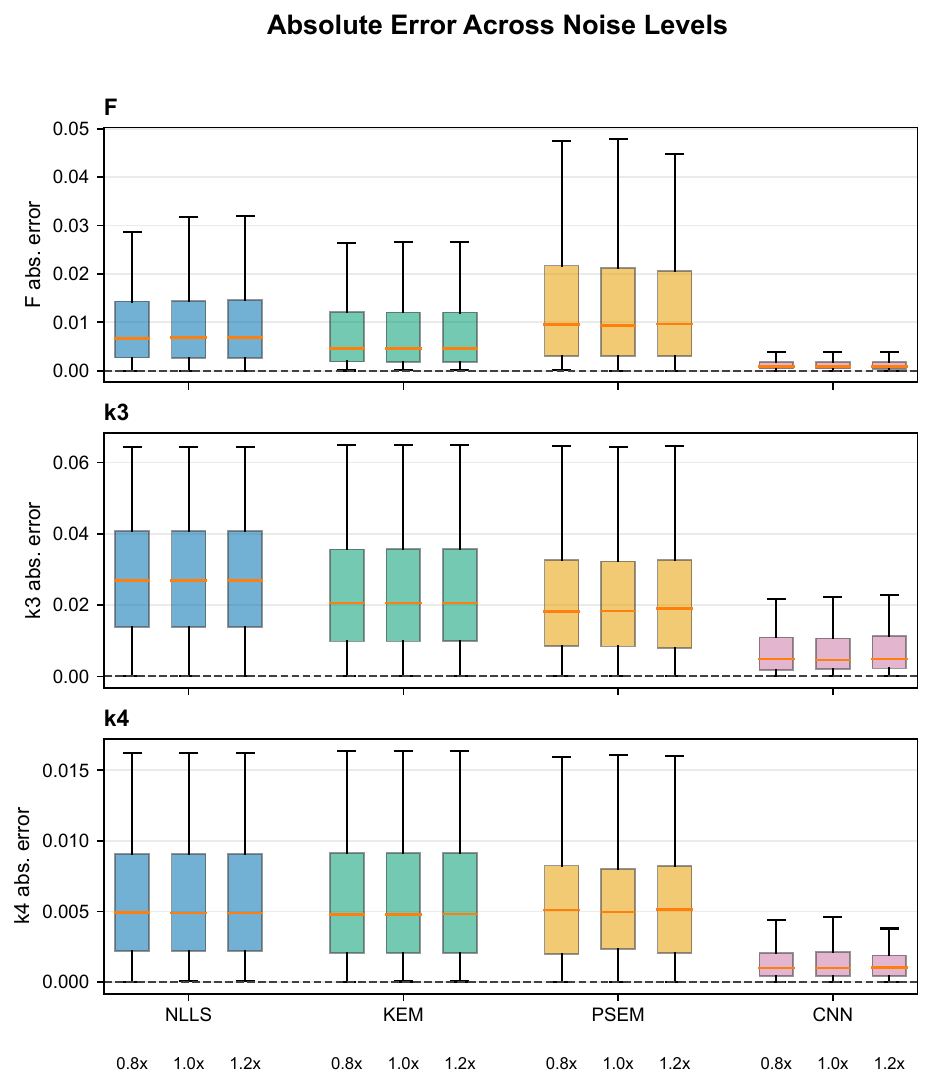}
    \caption{Absolute error distributions for the estimated parameters $F$, $k_3$, and $k_4$ across increasing noise levels (0.8$\times$, 1.0$\times$, and 1.2$\times$). For each parameter, boxplots summarize the absolute error for each estimation method (NLLS, KEM, PSEM, and CNN), grouped by noise level within method. Boxes indicate the interquartile range with the median shown, and whiskers denote the data range excluding outliers.}
    \label{fig:abs_err_noise}
\end{figure}

We regenerated the 200 test samples at 0.8$\times$ and 1.2$\times$ noise (default 1.0$\times$) using the same random seeds and input function. Results are shown in Fig.~\ref{fig:abs_err_noise}.

\subsection{NLLS initialization and convergence behavior}
\label{sec:nlls_init}

To assess the sensitivity of NLLS to initialization, we evaluated convergence behavior using multiple starting points within the feasible parameter space. Specifically, we selected three interior initializations within the 3D parameter cube defined by $(F, k_3, k_4)$, spanning low, medium, and high values within the allowable range (Table~\ref{tab:sim_params}) for 27 points.

\begin{figure}[t]
\centering
\includegraphics[width=.8\linewidth]{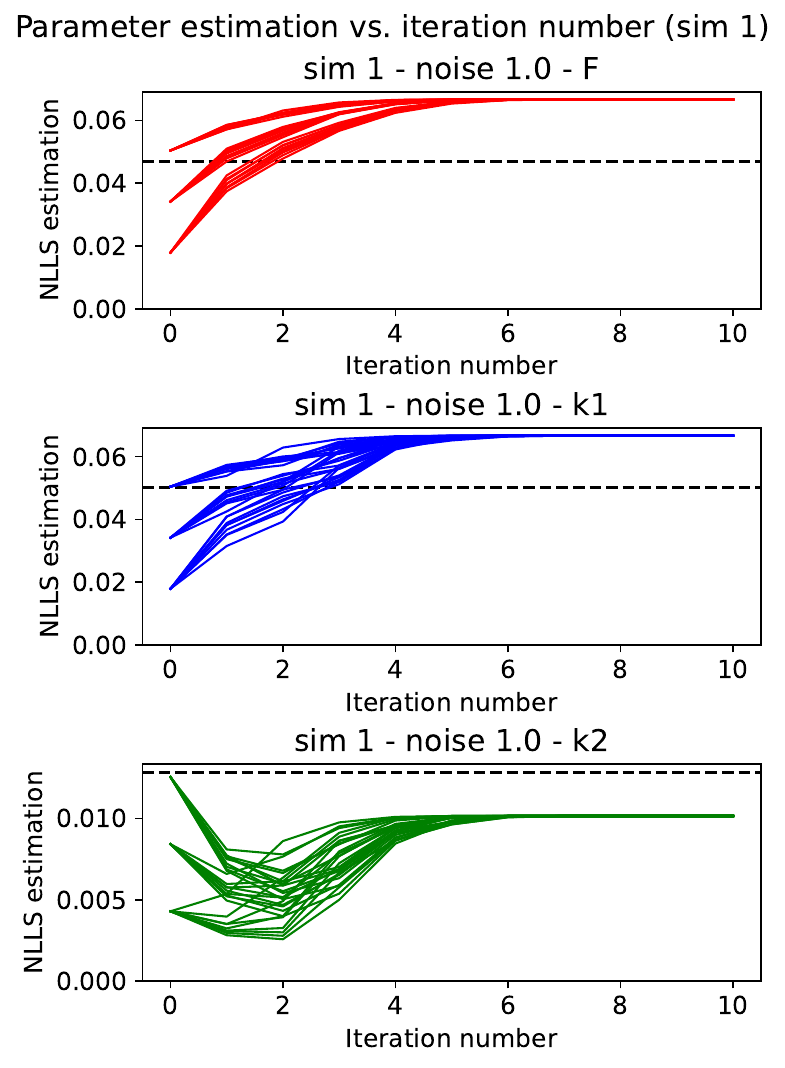}
\caption{
Parameter estimates as a function of iteration number for multiple initializations within the parameter space. Each subplot corresponds to one parameter ($F$, $k_3$, $k_4$). Different curves represent different starting points within the feasible region. All initializations converge to the same solution within a small number of iterations, indicating robustness to initialization under the evaluated conditions. The dashed line denotes the ground truth value.
}
\label{fig:nlls_iter}
\end{figure}

Figure~\ref{fig:nlls_iter} shows the initial and final parameter estimates for each initialization. For all three parameters ($F$, $k_3$, and $k_4$), the estimates converge to the same final value regardless of the starting point. This indicates that, under the evaluated conditions, the optimization is not sensitive to initialization.

\subsection{Performance with true $F_p, v$} 
\label{sec:true_other_params}
To assess the impact of nuisance-parameter misspecification, we repeated the analysis using the true values of $F_p$ and $v$ for the analytical methods. Table~\ref{tab:true_param} shows the mean relative error averaged across noise conditions (0.8x and 1.2x) for the 2~s framing protocol with the default input function over conditions where $F_p, v$ was set to population mean (introducing model misspecification) vs. $F_p, v$ set to true value.

When $F_p$ and $v$ are misspecified all analytical methods exhibit substantial error across parameters, particularly for $k_3$ and $k_4$. In contrast, when the true values are provided, NLLS shows a marked improvement, with errors reduced from 30.15\% to 1.50\% for $F$, 107.36\% to 14.74\% for $k_3$, and 118.41\% to 20.77\% for $k_4$. KEM shows moderate improvement for $F$ but remains sensitive to misspecification in $k_3$ and $k_4$, while PSEM shows limited improvement overall.

These results indicate that the apparent performance gap between CNN and analytical methods in the main experiments is driven in part by sensitivity to nuisance-parameter misspecification. When the correct $F_p$ and $v$ are provided, NLLS achieves the lowest error across all parameters, outperforming the CNN.
\begin{table}[t]
\centering
\caption{Mean relative error (\%) for 2 s frame and default input function over conditions where $F_p, v$ was set to population mean (model misspecification) vs. $F_p, v$ set to true value.}
\begin{tabular}{llcccc}
\toprule
\textbf{Setting} & \textbf{Param.} & \textbf{NLLS} & \textbf{KEM} & \textbf{PSEM} & \textbf{CNN} \\
\midrule
\multirow{3}{*}{Misspecified $F_p, v$}
 & $F$   & 30.15 & 29.19 & 57.92 & 10.85 \\
 & $k_3$ & 107.36 & 166.01 & 146.84 & 27.34 \\
 & $k_4$ & 118.41 & 121.37 & 137.07 & 35.24 \\
\midrule
\multirow{3}{*}{True $F_p, v$}
 & $F$   & 1.50  & 7.88  & 24.35 & 10.85 \\
 & $k_3$ & 14.74 & 163.48 & 148.79 & 27.34 \\
 & $k_4$ & 20.77 & 124.88 & 138.57 & 35.24 \\
\bottomrule
\end{tabular}
\label{tab:true_param}
\end{table}

\section*{References}
\addcontentsline{toc}{section}{\numberline{}References}
\vspace*{-20mm}

\bibliographystyle{medphy}   
\bibliography{pf.bib}      %

\end{document}